\documentstyle[preprint,aps]{revtex}

\begin{document}
\draft
\preprint{LBL-36594}

\title{Open Charm Production in an Equilibrating Parton Plasma
}

\author{Peter L\'evai$^a$, Berndt M\"uller$^b$, and Xin-Nian Wang$^c$}

\address{{$^a$Research Institute for Particle and Nuclear Physics,\\
    \baselineskip=12pt H-1525 Budapest 114, P. O. Box 49, Hungary}\\
  {$^b$Department of Physics, Duke University,\\
    \baselineskip=12pt Durham, NC 27708-0305}\\
  {$^c$Nuclear Science Division, MS 70A-3307, \\
    \baselineskip=12pt Lawrence Berkeley Laboratory, Berkeley, CA 94720}
}

\date{November 21, 1994}
\maketitle
\begin{abstract}

  Open charm production during the equilibration of a gluon
dominated parton plasma is calculated, with both the time-dependent
temperature and parton densities given by a set of rate equations.
Including pre-thermal production, the total enhancement of open
charm production over the initial gluon fusion depends sensitively
on the initial parton density and the effective temperature. The
dependence of the pre-thermal charm production on the space-momentum
correlation in the initial parton phase-space distribution is also
discussed.
\end{abstract}


\narrowtext

\section{Introduction}

At extremely high energies, nucleus-nucleus collisions may be
described by parton interactions in the framework of perturbative
QCD (pQCD)-inspired
models \cite{HIJING,PCM,DTUNUC}. In this framework, hard or semihard
scatterings among partons dominate the reaction dynamics. They can liberate
partons from the individual confining nucleons, thus producing large
amount of transverse energy in the central region \cite{JBAM,KLL},
and drive the initially produced parton system toward
equilibrium \cite{PCM,SHUR,BDMTW,XS93}. In principle,
the same kind of hard processes,  such as open charm
production \cite{BMXW92,KG93,LG94}, direct photon and dilepton
production \cite{SX93,STLD}, can also be used as direct probes of the early
parton dynamics and the evolution of the quark-gluon plasma.
Unlike strange quarks, charm quarks cannot be easily produced
during the mixed and hadronic phases of the dense matter
since the charm mass is much larger than the corresponding
temperature scale. The only period when charm quarks can
be easily produced is during the early stage of the parton
evolution when the effective temperature is still high. At this
stage, the parton gas is still not fully equilibrated yet so that
the temperature is only an effective parameter describing
the average momentum scale. By measuring this pre-equilibrium
charm production, one can thus probe the initial parton
density in phase space and shed light on the equilibration
time \cite{BMXW92}.

Roughly speaking, ultrarelativistic heavy-ion collisions in
a partonic picture can be divided into three stages: (1)
During the early stage, hard or semihard parton scatterings,
which happen on a time scale of about 0.2 fm/$c$, produce a hot
and dilute parton gas. This parton gas is dominated by gluons and
its quark content is far below the chemical equilibrium value.
Multiple hard scatterings suffered by a single
parton during this short period of time when the beam partons
pass through each other are suppressed due to the interference
embedded in the Glauber formula for multiple scatterings \cite{WANG95}.
This leads to the observed disapparence of the Cronin effect
at high energy and at large transverse momentum \cite{CRON2}.
Interference and parton fusion also lead to the depletion
of small-$x$ partons in the effective parton distributions
inside a nucleus \cite{BRLU,EQW94}. This nuclear shadowing
of parton distributions
reduces the initial parton production  \cite{WG92}.
(2) After two beams of partons pass through each other, the produced
parton gas in the central rapidity region starts its evolution
toward (kinetic) thermalization mainly through elastic
scatterings and expansion. The kinematic separation
of partons through free-streaming gives an estimate of the time scale
$\tau_{\rm iso}\sim 0.5 - 0.7$ fm/$c$ \cite{BDMTW,KEXW94a},
when local isotropy in momentum distributions is reached.
(3) Further evolution of the parton gas toward
a fully (chemically) equilibrated parton plasma is
dictated by the parton proliferation through induced
radiation and gluon fusion. Due to the consumption of
energy by the additional parton production, the effective temperature
of the parton plasma cools down considerably faster than the ideal
Bjorken's scaling solution. Therefore, the life time of the plasma
is reduced before the temperature drops below the
QCD phase transition temperature \cite{BDMTW}.

Similarly, charm production can also be divided into three
different contributions
in the history of the evolution of the parton system: (1) initial
production during the overlapping period ; (2) pre-thermal
production from secondary parton scatterings during the thermalization,
$\tau<\tau_{\rm iso}$; (3) and thermal production during the parton
equilibration, $\tau>\tau_{\rm iso}$, in the expanding system.
In this paper, we will first review the equilibration of the initially
produced parton gas in Sec. II, incorporating the result of an
improved perturbative QCD analysis of Landau-Pomeranchuk-Migdal (LPM)
effect \cite{GWLPM1,GWLPM2}. Then we will discuss the three stages of open
charm production in a reversed order, starting with the charm
production during the final stage of parton equilibration in
Sec. III. For pre-thermal charm production, we will consider
the space-momentum correlation in the initial parton
phase-space distributions, which will suppress open charm
production during this period as compared to previous
estimates \cite{BMXW92}. In Sec. IV, we will compare the results
to the charm production during the initial hard or semi-hard
scatterings and also to the results in Geiger's calculation \cite{KG93}
which is about 40-50 times higher than our estimates here.
We will also discuss the change in charm production due to the
uncertainties in the initial parton density and effective
temperature. Finally we give our conclusions in Sec. V.

\section{Parton Production and Equilibration}

At collider energies ($\sqrt{s}> 100$ GeV), hard or semihard
parton scatterings are believed to be the dominant mechanism
for transverse energy production in the central region \cite{JBAM,KLL}.
These hard processes happen on a short time scale and they
generally break color coherence inside individual nucleons \cite{BMW92}.
After the fast parton pass through each other and leave the
central region, a partonic gas will be left behind which is not
immediately in thermal and chemical equilibrium.
The partons inside such a system will then undergo further
interactions and free-streaming. Neglecting parton scatterings
in this period of time, the kinematic separation of partons with
different rapidities in a cell establishes local
momentum isotropy at the time of the order of $\tau_{\rm iso}=0.7$
fm/$c$ \cite{BDMTW,KEXW94a}. If we assume this is the actual kinetic
equilibration (or thermalization) time for the partonic system,
The subsequent chemical equilibration can then be described
by a set of rate equations. In this section we will review
parton equilibration following Ref.~\cite{BDMTW} with improved
estimate of the gluon equilibration rate.

\subsection{Initial conditions: a hot and dilute gluonic gas}

Currently there are many models for incorporating hard and semi-hard
processes in hadronic and nuclear collisions \cite{HIJING,PCM,DTUNUC}.
We will use the results of the HIJING Monte Carlo model \cite{HIJING}
to estimate the initial parton production. In this model,
multiple hard or semi-hard parton scatterings with initial
and final state radiation are combined together with Lund
string phenomenology \cite{LUND} for the accompanying soft
nonperturbative interactions.

Let us first estimate the initial conditions at time, $\tau_{\rm iso}$,
from the HIJING results.
Since we are here primarily interested in the chemical equilibration
of the parton gas which has already reached local isotropy in momentum
space, we shall assume that the parton distributions
can be approximated by thermal phase space distributions with
non-equilibrium fugacities $\lambda_i$:
\begin{equation}
f(k;T,\lambda_i)  = \lambda_i\left( e^{u\cdot k /T} \pm
\lambda_i\right)^{-1}, \label{eq:eq1}
\end{equation}
where $u^{\mu}$ is the
four-velocity of the local comoving reference frame.  When the
parton fugacities $\lambda_i$ are much less than unity as
may happen during the early evolution of the parton system,
we can neglect the quantum corrections in Eq.~(\ref{eq:eq1}) and
write the momentum distributions in the factorized form,
\begin{equation}
\label{19}
f(k;T,\lambda _i)=\lambda _i\left (e^{ u\cdot k /T}\pm 1\right)^{-1}.
\end{equation}
Using this form of distributions, one has the parton and energy
densities,
\begin{equation}
n = (\lambda_g a_1 +\lambda_q b_1)T^3, \quad \varepsilon
= (\lambda_g a_2 +\lambda_q b_2) T^4. \label{eq:eq2}
\end{equation}
where $a_1=16\zeta (3)/\pi^2\approx 1.95$, $a_2=8\pi^2/15\approx 5.26$,
for a Bose distribution, $b_1=9\zeta (3)N_f/\pi^2\approx 2.20$ and
$b_2=7\pi^2N_f/20 \approx 6.9$ for a Dirac distribution. For a baryon
symmetric system, $\lambda_q=\lambda_{\bar q}$. Since boost invariance
has been demonstrated to be a good approximation for
the initially produced partons \cite{KEXW94a}, we can then
estimate the initial parton fugacities, $\lambda_{g,q}^0$
and temperature $T_0$  from
\begin{equation}
n_0 = \frac{1}{\pi R^2_{A} \tau_{\rm iso}} \frac{dN}{dy}\; ,
\quad \varepsilon_0 = n_0 \frac{4}{\pi}\langle k_T\rangle, \label{eq:eq3}
\end{equation}
where $\langle k_T\rangle$ is the average transverse momentum.
The quark fugacity is taken as $\lambda_q^0 = 0.16 \lambda_g^0$,
corresponding to a ratio 0.14 of the initial quark(antiquark)
number to the total number of partons.  Table \ref{table1}
shows these relevant quantities at the moment $\tau_{\rm iso}$,
for Au + Au collisions at Brookhaven National Laboratory
Relativistic Heavy Ion Collider (RHIC) and CERN Large Hadron
Collider (LHC) energies. One can observe
that the initial parton gas is rather hot reflecting the large
average transverse momentum. However, the parton gas is very
dilute as compared to the ideal gas at the same temperature. The
gas is also  dominated by gluons with its quark content far below
the chemical equilibrium value. We should emphasize that the initial
conditions listed here result from HIJING
calculation of parton production through semihard scatterings.
Soft partons, {\em e.g.}, due to parton production from the color
field \cite{KEMG}, are not included.

\subsection{Master rate equations}

In general, chemical reactions among partons can be quite complicated
because of the possibility of initial and final-state gluon radiations.
Interference effects due to multiple scatterings inside a dense medium,
{\em i.e.}, LPM  suppression of soft gluon radiation has to be taken
into account. One lesson one has learned from LPM effect \cite{GWLPM1,GWLPM2}
is that the radiation between two successive scatterings is the
sum, {\em on the amplitude level}, of both the initial state
radiation from the first scattering and the final state radiation
from the second one. Since the off-shell parton is space-like
in the first amplitude and time-like in the second, the picture
of a time-like parton propagating inside a medium in the parton
cascade simulations \cite{PCM} shall break down. Instead,
we shall here consider both initial and final state radiations
together associated with a single scattering (To the leading order,
a single additional gluon is radiated, such as $gg\to ggg$),
in which we can include LPM effect by a radiation suppression
factor. The analysis of QCD LPM effect in Ref.~\cite{GWLPM1,GWLPM2}
has been done for a fast parton traveling inside a parton
plasma. We will use the results for radiations off thermal
partons who average energy is about $T$, since we expect
the same physics to happen.

In order to permit the approach to chemical equilibrium, the reverse
process, {\em i.e.}, gluon absorption, has to be included as well, which is
easily achieved making  use of detailed balance.
We consider only the dominant process $gg\to ggg$.
Radiative processes involving quarks have substantially smaller
cross sections in pQCD, and quarks are considerably less
abundant than gluons in the initial phase of the chemical evolution of
the parton gas.  Here we are interested in understanding the basic
mechanisms underlying the formation of a chemically equilibrated
quark-gluon plasma, and the essential time-scales.  We hence restrict
our considerations to the dominant reaction mechanism for the
equilibration of each parton flavor.  These are just four
processes \cite{MSM86}:
\begin{equation}
gg \leftrightarrow ggg, \quad gg\leftrightarrow
q\overline{q}.\label{eq:eq4}
\end{equation}
Other scattering processes ensure the maintenance of thermal
equilibrium $(gg\leftrightarrow gg, \; gq \leftrightarrow gq$, etc.)
or yield corrections to the dominant reaction rates
$(gq\leftrightarrow qgg$, etc.).

Restricting to the reactions in Eq.~(\ref{eq:eq4}) and assuming
that elastic parton scatterings are sufficiently rapid to maintain
local thermal equilibrium, the evolution of the parton densities
is governed by the master equations \cite{BDMTW}:
\begin{eqnarray}
\partial_{\mu}(n_gu^{\mu}) &= &
 \frac{1}{2}\sigma_3 n_g^2 \left( 1-\frac{n_g}{\tilde n_g}\right)
 -\frac{1}{2}\sigma_2 n_g^2 \left( 1 - \frac{n_q^2 \tilde n_g^2}
 {\tilde n_q^2 n_g^2}\right), \label{eq:eq5}\\
 \partial_{\mu} (n_qu^{\mu}) &= & \frac{1}{2}\sigma_2 n_g^2
 \left( 1 - \frac{n_q^2 \tilde n_g^2}
 {\tilde n_q^2 n_g^2}\right), \label{eq:eq6}
\end{eqnarray}
where ${\tilde n_i}\equiv n_i/\lambda_i$ denote the densities
with unit fugacities, $\lambda_i=1$, $\sigma_3$ and $\sigma_2$
are thermally averaged, velocity weighted cross sections,
\begin{equation}
\sigma_3 = \langle\sigma(gg\to ggg)v\rangle, \quad \sigma_2 =
\langle \sigma (gg\to q\bar q)v\rangle. \label{eq:eq7}
\end{equation}
We have also assumed detailed balance and a baryon symmetric
matter, $n_q=n_{\bar q}$. If we neglect effects of viscosity
due to elastic scattering \cite{KEMG,VISC}, we then have the
hydrodynamic equation
\begin{equation}
\partial_{\mu} (\varepsilon u^{\mu}) + P\;\partial_{\mu} u^{\mu} = 0,
\label{eq:eq8}
\end{equation}
which determines the evolution of the energy density.

For a time scale $\tau\ll R_A$, we can neglect the transverse
expansion and consider the expansion of the parton plasma
purely longitudinal, which leads to the Bjorken's scaling
solution \cite{BJOR} of the hydrodynamic equation:
\begin{equation}
{d\varepsilon\over d\tau} + {\varepsilon+P\over\tau} = 0. \label{eq:eq9}
\end{equation}

We further assume the ultrarelativistic equation of motion,
$\varepsilon=3 P$ with $n_i$ and $\varepsilon$ given by Eq.~(\ref{eq:eq2}).
We can then solve the hydrodynamic equation,
\begin{equation}
  [\lambda_g + \frac{b_2}{a_2}\lambda_q]^{3/4} T^3\tau = \hbox{const.} \;\; ,
  \label{eq:eq10}
\end{equation}
and rewrite the rate equation in terms of time evolution of the
parameters $T(\tau)$, $\lambda_g(\tau)$ and $\lambda_q(\tau)$,
\begin{eqnarray}
\frac{\dot\lambda_g}{\lambda_g} + 3\frac{\dot T}{T} + \frac{1}{\tau} &=
&R_3 (1-\lambda_g)-2R_2 \left(1- \frac{\lambda_q^2}{\lambda_g^2} \right)
        \label{eq:eq11} \\
\frac{\dot\lambda_q}{\lambda_q} + 3\frac{\dot T}{T} + \frac{1}{\tau} &=
&R_2 {a_1\over b_1} \left( \frac{\lambda_g}{\lambda_q} -
\frac{\lambda_q}{\lambda_g}\right), \label{eq:eq12}
\end{eqnarray}
where the density weighted reaction rates $R_3$ and $R_2$ are defined as
\begin{equation}
R_3 = \textstyle{{1\over 2}} \sigma_3 n_g, \quad
R_2 = \textstyle{{1\over 2}} \sigma_2 n_g.  \label{eq:eq13}
\end{equation}
Notice that for a fully equilibrated system ($\lambda_g=\lambda_q=1$),
Eq. (\ref{eq:eq10}) corresponds to the Bjorken solution,
$T(\tau)/T_0=(\tau_0/\tau)^{1/3}$.

\subsection{Parton equilibration rates}

To take into account of the LPM effect in the calculation of
the reaction rate $R_3$ for $gg\rightarrow ggg$,
we simply impose the LPM suppression of
the gluon radiation whose effective formation time $\tau_{\rm QCD}$
is much longer than the mean-free-path $\lambda_f$ of multiple
scatterings to each $gg\rightarrow ggg$ process.
In the mean time, the LPM effect also regularizes
the infrared divergency associated with QCD radiation. However,
$\sigma_3$ still contains infrared singularities in the gluon
propagators. For an equilibrium system one can in principle apply
the resummation technique developed by Braaten and Pisarski \cite{BP90}
to regularize the electric part of the propagators, though the
magnetic sector still has to be determined by an unknown magnetic
screening mass which can only be calculated nonperturbatively \cite{TBBM93}
up to now. Since we are dealing with a nonequilibrium system,
Braaten and Pisarski's resummation may not be well defined. As an
approximation, we will use the Debye screening mass \cite{BMW92},
\begin{equation}
\mu_D^2 = {6g^2\over \pi^2} \int_0^{\infty} kf(k) dk
=4\pi\alpha_s T^2\lambda_g, \label{eq:eq14}
\end{equation}
to regularize all singularities in both the scattering cross
sections and the radiation amplitude.

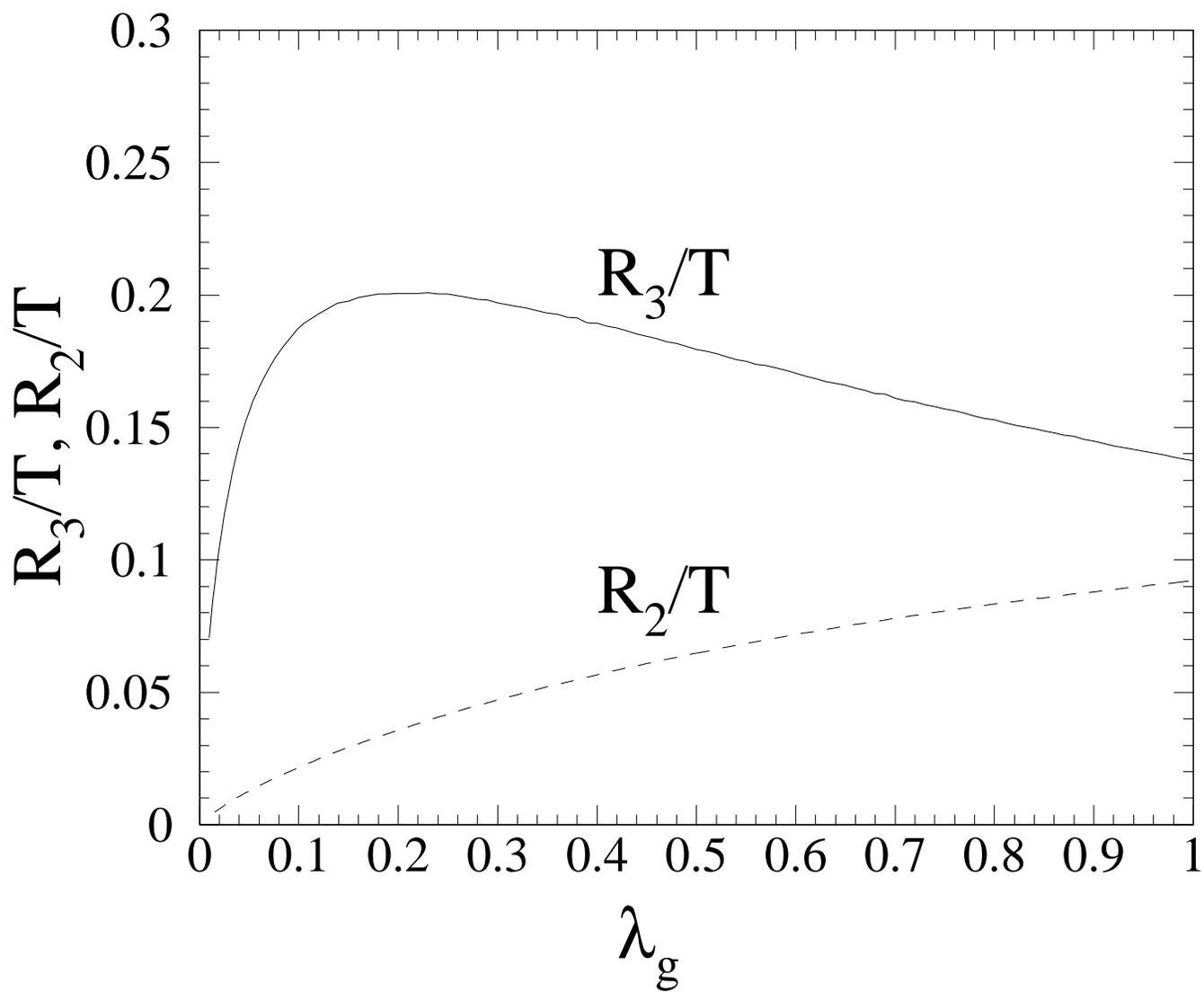
\begin{figure}
\caption{The scaled gluon production rate $R_3/T$ (solid line) for
         $gg\rightarrow ggg$ and the quark production rate
         $R_2/T$ (dashed line) for $gg\rightarrow q\bar{q}$
         are shown as function of the gluon fugacity $\lambda_g$ for
         $\alpha_s$ = 0.3.}
\label{fig1}
\end{figure}

To further simplify the calculation we approximate the
LPM suppression factor in Ref.~\cite{GWLPM1,GWLPM2} by a $\theta$-function,
$\theta(\lambda_f-\tau_{\rm QCD})$, where
\begin{equation}
  \tau_{\rm QCD}=\frac{C_A}{2C_2}\frac{2\cosh y}{k_{\perp}},
\end{equation}
is the effective formation time of the gluon radiation in QCD
which depends on the second Casimir of the beam parton representation
in $SU(3)$, {\em e.g.}, $C_2=C_A=3$ for a gluon. In the previous
calculation of the interaction rate \cite{BDMTW}, this color factor was
not taken into account. The modified
differential cross section for $gg\rightarrow ggg$ is then,
\begin{equation}
  \frac{d\sigma_3}{dq_{\perp}^2 dy d^2k_{\perp}}
  =\frac{d\sigma_{\rm el}^{gg}}{dq_{\perp}^2}\frac{dn_g}{dy d^2k_{\perp}}
  \theta(\lambda_f-\tau_{QCD})\theta(\sqrt{s}-k_{\perp}\cosh y),
\end{equation}
where the second
step-function accounts for energy conservation, and
$s=18T^2$ is the average squared center-of-mass energy of two
gluons in the thermal gas. The regularized gluon density distribution
induced by a single scattering is \cite{GUNION},
\begin{equation}
  \frac{dn_g}{dy d^2k_{\perp}} =\frac{C_A\alpha_s}{\pi^2}
  \frac{q_{\perp}^2}{k_{\perp}^2[({\bf k}_{\perp}
    -{\bf q}_{\perp})^2 +\mu_D^2]}. \label{eq:dng}
\end{equation}
Similarly, the regularized small angle $gg$ scattering cross is,
\begin{equation}
  \frac{d\sigma_{\rm el}^{gg}}{dq_{\perp}^2}
  =\frac{9}{4}\frac{2\pi\alpha_s^2}{(q_{\perp}^2+\mu_D^2)^2}.
\end{equation}
The mean-free-path for elastic scatterings is then,
\begin{equation}
  \lambda_f^{-1}\equiv n_g\int_0^{s/4}dq_{\perp}^2
  \frac{d\sigma_{\rm el}^{gg}}{dq_{\perp}^2}
  =\frac{9}{8}a_1\alpha_s T\frac{1}{1+8\pi\alpha_s\lambda_g/9},
\end{equation}
which depends very weekly on the gluon fugacity $\lambda_g$
as compared to the independent one used in a previous
study \cite{BDMTW}. Using,
\begin{equation}
  \int_0^{2\pi}d\phi \frac{1}{({\bf k}_{\perp}-{\bf q}_{\perp})^2+\mu_D^2}
    =\frac{2\pi}{\sqrt{(k_{\perp}^2+q_{\perp}^2+\mu_D^2)^2
        -4q_{\perp}^2k_{\perp}^2}},
\end{equation}
we can complete part of the integrations and have,
\begin{equation}
  R_3/T=\frac{32}{3a_1}\alpha_s\lambda_g(1+8\pi\alpha_s\lambda_g/9)^2
  {\cal I}(\lambda_g),
\end{equation}
where ${\cal I}(\lambda_g)$ is a function of $\lambda_g$,
\begin{eqnarray}
{\cal I}(\lambda_g)=\int_1^{\sqrt{s}\lambda_f}dx
\int_0^{s/4\mu_D^2}&dz& \frac{z}{(1+z)^2}
  \left\{ {\cosh^{-1}(\sqrt{x}) \over
  x\sqrt{[x+(1+z)x_D]^2-4x\;z\;x_D}}\right. \nonumber \\
  &+& \left. \frac{1}{s\lambda_f^2}{\cosh^{-1}(\sqrt{x}) \over
  \sqrt{[1+x(1+z)y_D]^2-4x\;z\;y_D}}\right\},
\end{eqnarray}
where $x_D=\mu_D^2\lambda_f^2$ and $y_D=\mu_D^2/s$.
We can evaluate the integration numerically and find out
the dependence of $R_3/T$ on the gluon fugacity $\lambda_g$.
In Fig.~\ref{fig1}, $R_3/T$ is plotted versus $\lambda_g$
for a coupling constant $\alpha_s=0.3$. The gluon production
rate increases with $\lambda_g$ and then saturates when
the system is in equilibrium.
Note that in principle one should multiply the phase-space
integral by $1/3!$ to take into account of the symmetrization
of identical particles in $gg \rightarrow ggg$ as in Ref.~\cite{SX93}.
However, for the dominant soft radiation we consider here,
the radiated soft gluon does not overlap with the two incident
gluons in the phase-space. Thus we only multiply the cross section
by $1/2!$ to obtain the equilibration rate.

The calculation of the quark equilibration rate $R_2$ for
$gg\rightarrow q\bar{q}$ is more straightforward.
Estimate in Ref.~\cite{BDMTW} gives,
\begin{equation}
R_2 = {1\over 2}\sigma_2 n_g \approx 0.24 N_f \;\alpha_s^2 \lambda_g T
\ln (5.5/\lambda_g). \label{56}
\end{equation}
The dashed line in Fig.~\ref{fig1} shows the normalized
rate $R_2/T$ for $N_f=2.5$, taking into account the reduced
phase space of strange quarks at moderate temperatures, as a
function of the gluon fugacity.

\subsection{Evolution of the parton plasma}

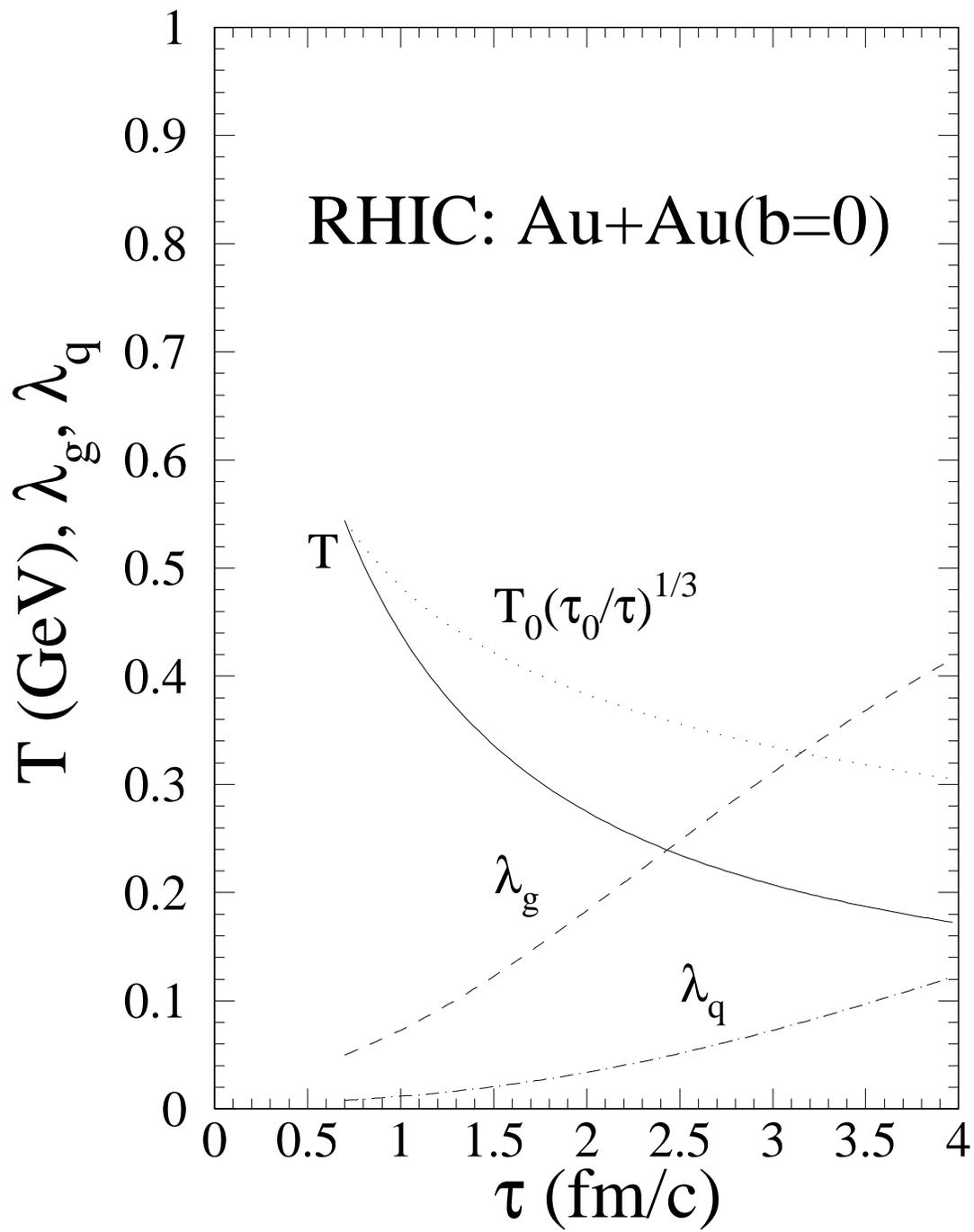
\begin{figure}
\caption{Time evolution of the temperature $T$ and the
         fugacities $\lambda_g$ and $\lambda_q$ of gluons and quarks in the
         parton plasma created in Au + Au collisions at the RHIC energy
         of $\protect\sqrt{s}=200$ AGeV. The initial values for
         $T,\; \lambda_g$ and $\lambda_q$ are determined from
         HIJING simulations and are listed in Table~\protect\ref{table1}.}
\label{fig2}
\end{figure}

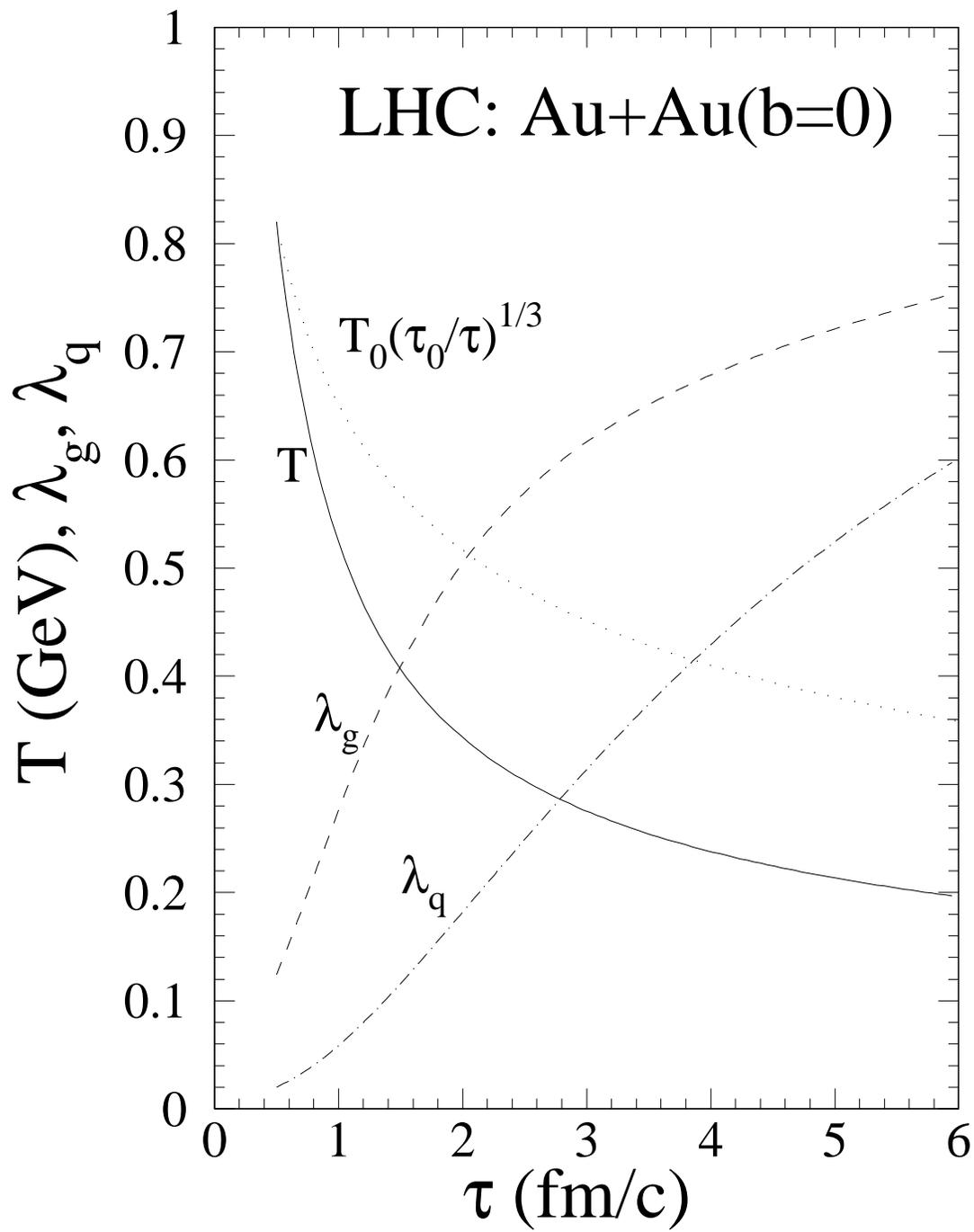
\begin{figure}
\caption{The same as in Fig.~\protect\ref{fig2}, except for LHC energy,
         $\protect\sqrt{s}=5.5 $ ATeV. }
\label{fig3}
\end{figure}

With the parton equilibration rates which in turn depend
on the parton fugacity, we can solve the master equations
self-consistently and obtain the time evolution of the temperature
and the fugacities. Shown in Figs.~\ref{fig2} and \ref{fig3}
are the time dependence of $T$, $\lambda_g$, and $\lambda_q$
for initial conditions listed in Table~\ref{table1} at RHIC and
LHC energies. We find that the parton gas cools considerably
faster than predicted by Bjorken's scaling solution
$(T^3\tau$ = const.) shown as dotted lines, because
the production of additional partons
approaching the chemical equilibrium state consumes an appreciable
amount of energy. The accelerated cooling, in turn, slows down the
chemical equilibration process, which is more apparent at RHIC
than at LHC energies. Therefore, the parton system can hardly
reach its equilibrium state before the effective temperature
drops below $T_c \approx 200$ MeV in a short period of time of
1-2 fm/$c$ at RHIC energy. At LHC energy, however, the parton
gas becomes very close to its equilibrium and the plasma
may exist in a deconfined phase for as long as 4-5 fm/$c$.
Another important observation is that quarks never
approach to chemical equilibrium at both energies.
This is partially due to the small initial quark fugacity
and  partially due to the small quark equilibration rate.

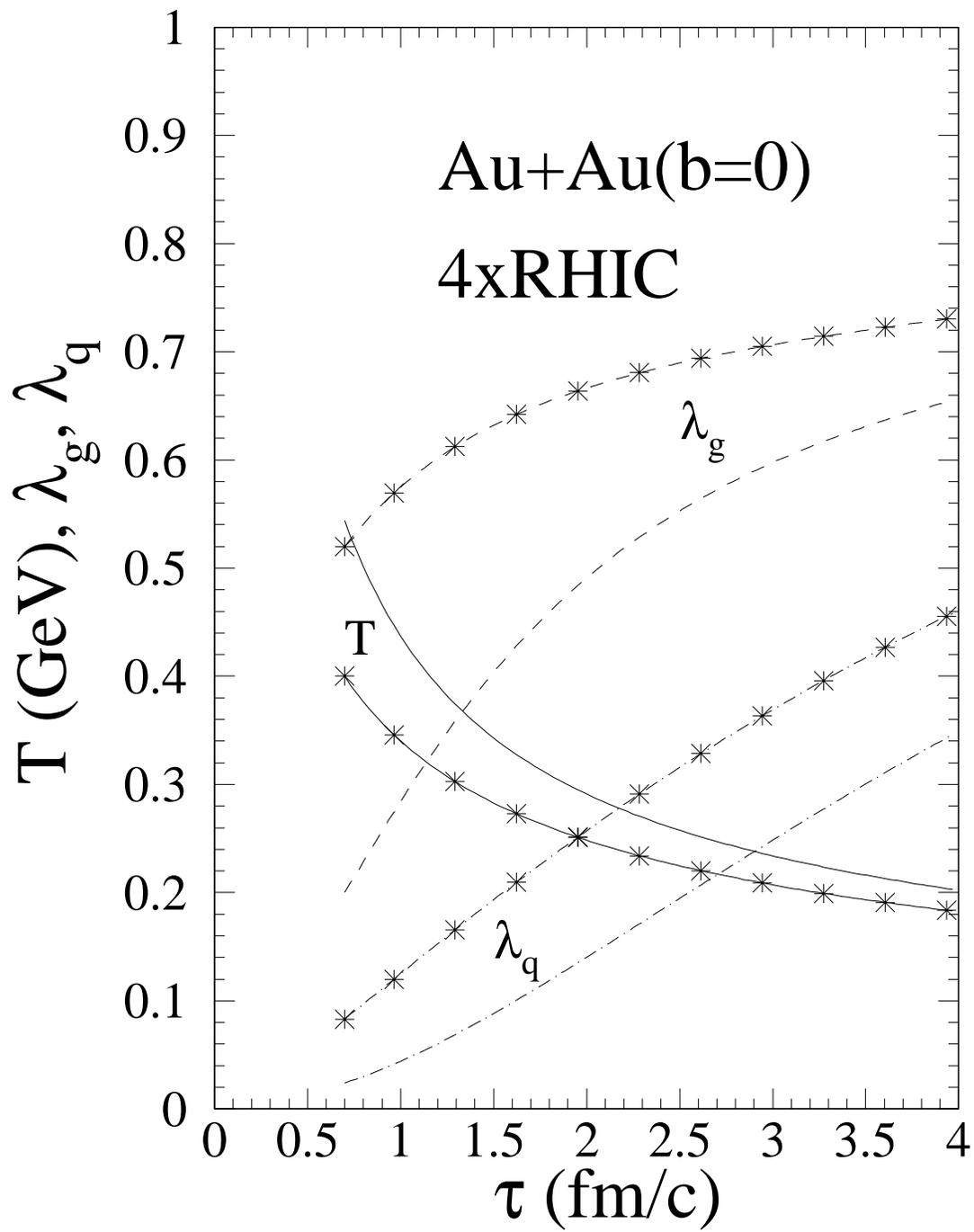
\begin{figure}
\caption{The same as in Fig.~\protect\ref{fig2}, except that the initial
         parton densities are 4 times higher than given in
         Table~\protect\ref{table1} with the same (ordinary lines),
         or reduced initial temperature, $T_0=0.4$ GeV (lines with stars)}
\label{fig4}
\end{figure}

We note that the initial conditions used here result
from the HIJING model calculation in which only initial direct parton
scatterings are taken into account.  Due to the fact that HIJING is
a pQCD motivated phenomenological model, there are some uncertainties
related to the initial parton production, as listed in Ref.~\cite{BDMTW}.
We can estimate the effect of the uncertainties in the
initial conditions on the parton gas evolution by multiplying
the initial energy and parton number densities at RHIC energy
by a factor of 4. This will result in the initial fugacities,
$\lambda_g^0=0.2$ and $\lambda_q^0=0.024$. With these high
initial densities, the parton plasma can evolve into a
nearly equilibrated gluon gas as shown in Fig.~\ref{fig4}.
The deconfined phase will also last longer for about 4 fm/$c$.
Though, the system is still dominated by gluons with few quarks
and antiquarks as compared to a fully chemical equilibrated
system. If the uncertainties in the initial conditions
are caused by the soft parton production from the color
mean fields, the initial effective temperature will decrease.
Therefore, we can alternatively increase the initial
parton density by a factor of 4 and decrease $T_0$ to
0.4 GeV at the same time. This leads to higher initial fugacities,
$\lambda_g^0=0.52$ and $\lambda_q^0=0.083$. As shown in
Fig.~\ref{fig4} by the curves with stars, this system evolves
faster toward equilibrium, however, with shorter life-time
in the deconfined phase due to the reduced initial
temperature.

We thus can conclude that perturbative parton
production and scatterings are very likely to produce
a quark-gluon plasma ( or more specifically a gluon plasma)
in ultrarelativistic heavy ion collisions at LHC energy.
However, the fate of the quark-gluon plasma at RHIC energy
has to be determined by a more careful examination of the
uncertainties in the initial conditions. These uncertainties
will surely affect the open charm production during the
equilibration as we shall discuss.

\section{Thermal charm production during equilibration}

With the given evolution of the parton gas, we can now
calculate open charm production during the parton equilibration.
Similar to light quarks, charm quarks are
produced through gluon fusion $gg\rightarrow c\bar{c}$ and
quark antiquark annihilation $q\bar{q}\rightarrow c\bar{c}$
during the evolution of the parton plasma. However, since
the number of charm quarks is very small as compared
to gluons and light quarks, we can neglect the back reactions,
$c\bar{c}\rightarrow gg,\ q\bar{q}$ and their effect on the parton
evolution. Given the phase-space density of the equilibrating
partons, $f_i(k)$, the differential production rate is \cite{BMXW92},
\begin{eqnarray}
  E\frac{d^3A}{d^3p}&=&\frac{1}{16(2\pi)^8}\int \frac{d^3k_1}{\omega_1}
  \frac{d^3k_2}{\omega_2}\frac{d^3p_2}{E_2}\delta^{(4)}(k_1+k_2-p-p_2)
  \nonumber \\
  & & \left[\frac{1}{2}g_G^2f_g(k_1)f_g(k_2)
  |\overline{\cal M}_{gg\rightarrow c\bar{c}}|^2+g_q^2f_q(k_1)f_{\bar{q}}(k_2)
  |\overline{\cal M}_{q\bar{q}\rightarrow c\bar{c}}|^2\right]
  , \label{eq:therm1}
\end{eqnarray}
where $g_G$=16, $g_q=6N_f$, are the degeneracy factors for gluons and
quarks (antiquarks) respectively,
$|\overline{\cal M}_{gg\rightarrow c\bar{c}}|^2$,
$|\overline{\cal M}_{q\bar{q}\rightarrow c\bar{c}}|^2$ are
the {\em averaged} matrix elements for $gg\rightarrow c\bar{c}$
and $q\bar{q}\rightarrow c\bar{c}$ processes, respectively,
\begin{eqnarray}
  \frac{|\overline{\cal M}_{gg\rightarrow c\bar{c}}|^2}{\pi^2\alpha_s^2} &=&
  \frac{12}{\hat{s}^2}(M^2-\hat{t})(M^2-\hat{u})+\frac{8}{3}
  \left(\frac{M^2-\hat{u}}{M^2-\hat{t}}
  +\frac{M^2-\hat{t}}{M^2-\hat{u}}\right) \nonumber \\
  &-&\frac{16M^2}{3} \left[ \frac{M^2+\hat{t}}{(M^2-\hat{t})^2}
+\frac{M^2+\hat{u}}{(M^2-\hat{u})^2} \right]
  -\frac{6}{\hat{s}}(2M^2-\hat{t}-\hat{u}) \nonumber \\
  &+&\frac{6}{\hat{s}}\frac{M^2(\hat{t}-\hat{u})^2}
  {(M^2-\hat{t})(M^2-\hat{u})}
  -\frac{2}{3}\frac{M^2(\hat{s}-4M^2)}{(M^2-\hat{t})(M^2-\hat{u})},\\
\frac{|\overline{\cal M}_{q\bar{q}\rightarrow c\bar{c}}|^2}{\pi^2\alpha_s^2}
&=&\frac{64}{9\hat{s}^2}
  \left[(M^2-\hat{t})^2+(M^2-\hat{u})^2+2M^2\hat{s}\right],
\end{eqnarray}
Due to the small charm density, we can neglect the Pauli blocking
of the final charm quarks. For large charm quark mass, $M$, we
can approximate the phase-space density $f_i(k)$ by a Boltzmann
distribution. We further assume that the distributions are
boost invariant, {\em i.e.},
\begin{equation}
  f_i(k)\approx \lambda_i e^{-k_{\perp}\cosh(y-\eta)},
\end{equation}
where $\eta=0.5\ln(t+z)/(t-z)$ is the spatial rapidity of a
space-time cell at $(t,z)$. Neglecting the transverse
expansion, the above assumption implies that the
space-time cell at $(t,z)$ have a flow velocity, $u=(\cosh\eta,\sinh\eta)$.
We can now complete the integral over $\eta$ in
$\int d^4x=\pi R_A^2\int d\eta d\tau$ and obtain,
\begin{eqnarray}
  \frac{dN_{\rm th}}{dyd^2p_{\perp}}&=&\frac{\pi R_A^2}{16(2\pi)^8}
  \int_{\tau_{\rm iso}}^{\tau_c}\tau d\tau \int p_{\perp 2}dp_{\perp 2}
  d\phi_2 dy_2 d\phi_{k1} dy_{k1} \frac{2k_{\perp 1}^2}{\hat{s}}
  2 K_0(Q_{\perp}/T) \nonumber \\
& &\left[\frac{1}{2}g_G^2\lambda_g^2
|\overline{\cal M}_{gg\rightarrow c\bar{c}}|^2
+g_q^2\lambda_q^2|\overline{\cal M}_{q\bar{q}\rightarrow c\bar{c}}|^2\right],
\label{eq:therm2}
\end{eqnarray}
where $K_0$ is the modified Bassel function and $\tau_c$ is the
time when the temperature, $T$, drops below 200 MeV.
The kinematic variables are chosen such that,
\begin{eqnarray}
  p_2&=&(M_{\perp 2}\cosh y_2,p_{\perp 2}\cos\phi_2,p_{\perp 2}\sin\phi_2,
  M_{\perp 2}\sinh y_2),
  \;\; M_{\perp 2}=\sqrt{M^2+p_{\perp 2}^2}; \nonumber \\
  k_i&=&k_{\perp i}(\cosh y_{ki},\cos\phi_{ki},\sin\phi_{ki},\sinh y_{ki}),
 \ \ i=1,2 \ .
\end{eqnarray}
The center-of-mass momentum,
$Q=(Q_{\perp}\cosh y_Q,{\bf q_{\perp}},Q_{\perp}\sinh y_Q)$,
is defined as $Q=p+p_2=k_1+k_2$, and
\begin{eqnarray}
  Q^2&=&\hat{s}=2[M^2+M_{T}M_{\perp 2}\cosh(y-y_2)
  -p_{\perp}p_{\perp 2}\cos\phi_2],\nonumber\\
  q_{\perp}^2&=&p_{\perp}^2+p_{\perp 2}^2
  +2p_{\perp}p_{\perp 2}\cos\phi_2,\nonumber \\
  Q_{\perp}^2&=&Q^2+q_{\perp}^2=M_{T}^2+M_{\perp 2}^2
  +2M_{T}M_{\perp 2}\cosh(y-y_2).
\end{eqnarray}
Using these variables and the energy-momentum conservation, we have,
\begin{eqnarray}
  k_{\perp 1}&=&\frac{Q^2/2}{M_{\perp}\cosh(y-y_{k1})
    +M_{\perp 2}\cosh(y_2-y_{k1})
    -q_{\perp}\cos\phi_{1q}}, \nonumber \\
  \cos\phi_{1q}&=&[p_{\perp}\cos\phi_{k1}
  +p_{\perp 2}\cos(\phi_2-\phi_{k1})]/q_{\perp}.
\end{eqnarray}
In the integral over $\tau$, we shall use the time evolution of
the temperature, $T(\tau)$, and fugacities, $\lambda_i(\tau)$,
as given in the previous section.

\section{Pre-thermal charm production}

Before the parton distributions reach local isotropy in momentum
space so that the rate equations can be applied to describe
the equilibration of the parton system, scatterings among
free-streaming partons can also lead to charm production. Since
the system during this period consists dominantly of gluons,
we shall only consider gluon fusions. To model the phase-space
distribution, we take into account the distribution of the
initial production points which spread over a region
of width,
\begin{equation}
  \Delta_k\approx \frac{2}{k_{\perp}\cosh y},
\end{equation}
in $z$ coordinate. Following Ref.~\cite{LG94}, we assume
free-streaming until $\tau_{\rm iso}$ and neglect the expansion in
the transverse direction. The correlated phase-space distribution
function is given by
\begin{equation}
  f(k,x)=\frac{1}{g_G\pi R_A^2}g(k_{\perp},y)
  \frac{e^{-(z-t\tanh y)^2/2\Delta_k^2}}{\sqrt{2\pi}\Delta_k}
  \theta(R_A-r)\theta(\tau_{\rm iso}\cosh y-t),
  \label{eq:phase}
\end{equation}
where $g(k_{\perp},y)$ is the parametrization of the parton spectrum
given by HIJING simulations,
\begin{equation}
  g(k_{\perp},y)=\frac{(2\pi)^3}{k}\frac{dN_g}{dyd^2k_{\perp}}
  =\frac{(2\pi)^2}{k}h(k_{\perp})\frac{1}{2Y}\theta(Y^2-y^2).
\end{equation}
The phase-space distribution is normalized such that
$\lim_{t \rightarrow \infty} g_G \int d^3 x f(k,x)/(2\pi)^3 =d^3N_g/d^3k$.
The function $h(k_{\perp})$
and the rapidity width $Y$ are given in Table~\ref{table2} for
central $Au+Au$ collisions at RHIC and LHC energies which
also gives the initial conditions as listed in Table~\ref{table1}.

Substituting the phase-space distribution, into Eq.~(\ref{eq:therm1}),
and integrate over space and time, we obtain the charm production
distribution in the pre-thermal period,
\begin{eqnarray}
  \frac{dN_{\rm pre}}{dyd^2p_{\perp}}&=&\frac{1}{16(2\pi)^8\pi R_A^2}
  \int p_{\perp 2}dp_{\perp 2} d\phi_2 dy_2 d\phi_{k1} dy_{k1}
  \frac{2k_{\perp 1}^2}{\hat{s}}
  g(k_{\perp 1},y_{k1})g(k_{\perp 2},y_{k2})\nonumber \\
  & &\frac{1}{2}|\overline{\cal M}_{gg\rightarrow c\bar{c}}|^2
  \frac{1}{\sqrt{2\pi}\Delta_{\rm tot}} \int_0^{t_f}dt
  e^{-t^2(\tanh y_1-\tanh y_2)^2/2\Delta^2_{\rm tot}},
\label{eq:preth}
\end{eqnarray}
\begin{equation}
  \Delta_{\rm tot}=\sqrt{\Delta_{k1}^2 + \Delta_{k2}^2},
  \;\; t_{\rm f}=\tau_{\rm iso}\min(\cosh y_{k1},\cosh y_{k2}).
\end{equation}
where the kinematic variables are similarly defined as in
Eq.~(\ref{eq:therm2}), and in addition,
\begin{eqnarray}
  k_{\perp 2}^2&=&Q_{\perp}^2+k_{\perp 1}^2
  -2k_{\perp 1}[M_{\perp}\cosh(y-y_{k1})
  +M_{\perp 2}\cosh(y_2-y_{k1})], \nonumber \\
  \sinh y_{k2}&=&[M_{\perp}\sinh y
  +M_{\perp 2}\sinh y_2-k_{\perp 1}\sinh y_{k1}]/k_{\perp 2}.
\end{eqnarray}

Note that the correlation between momentum and space-time in the
phase-space distribution was not considered in a previous
calculation \cite{BMXW92}. As we will show this correlation is
very important and will reduce the pre-thermal charm production
as compared to the uncorrelated
distributions. Similar effect was recently discussed by
Lin and Gyulassy in Ref.~\cite{LG94}, where formation time effect
is also included which is expected to further suppress pre-thermal charm
production.

\section{Initial fusion}

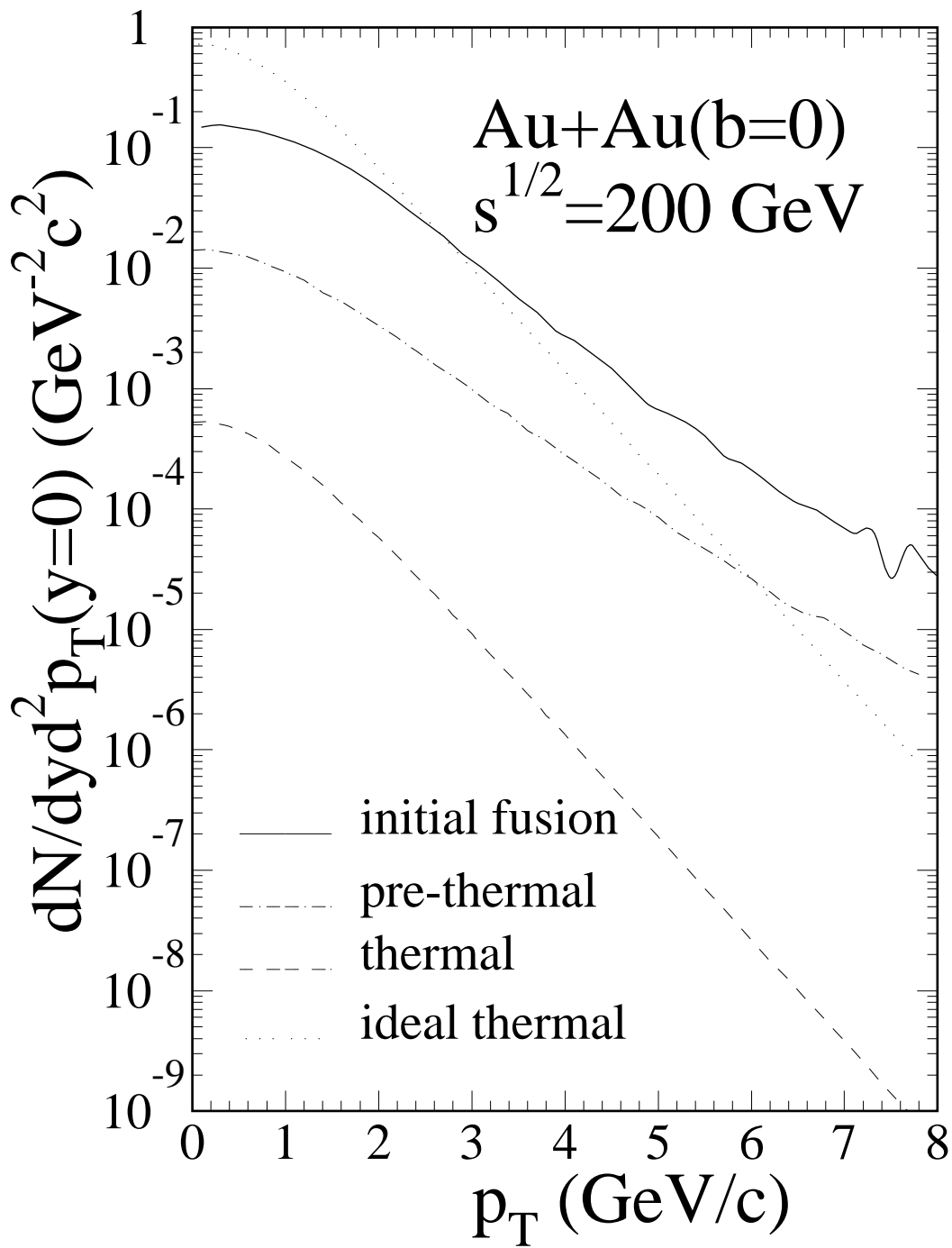
\begin{figure}
\caption{The $p_{\perp}$ distributions of initial (solid),
         prethermal (dot-dashed),
         and thermal (dashed) charm production for central $Au+Au$ collisions
         at RHIC energy, $\protect\sqrt{s}=200$ AGeV with initial
         conditions given in Tables \protect\ref{table1} and
         \protect\ref{table2}. The dotted line is the
         thermal production assuming an initial fully equilibrated QGP
         at the same temperature.}
\label{fig5}
\end{figure}

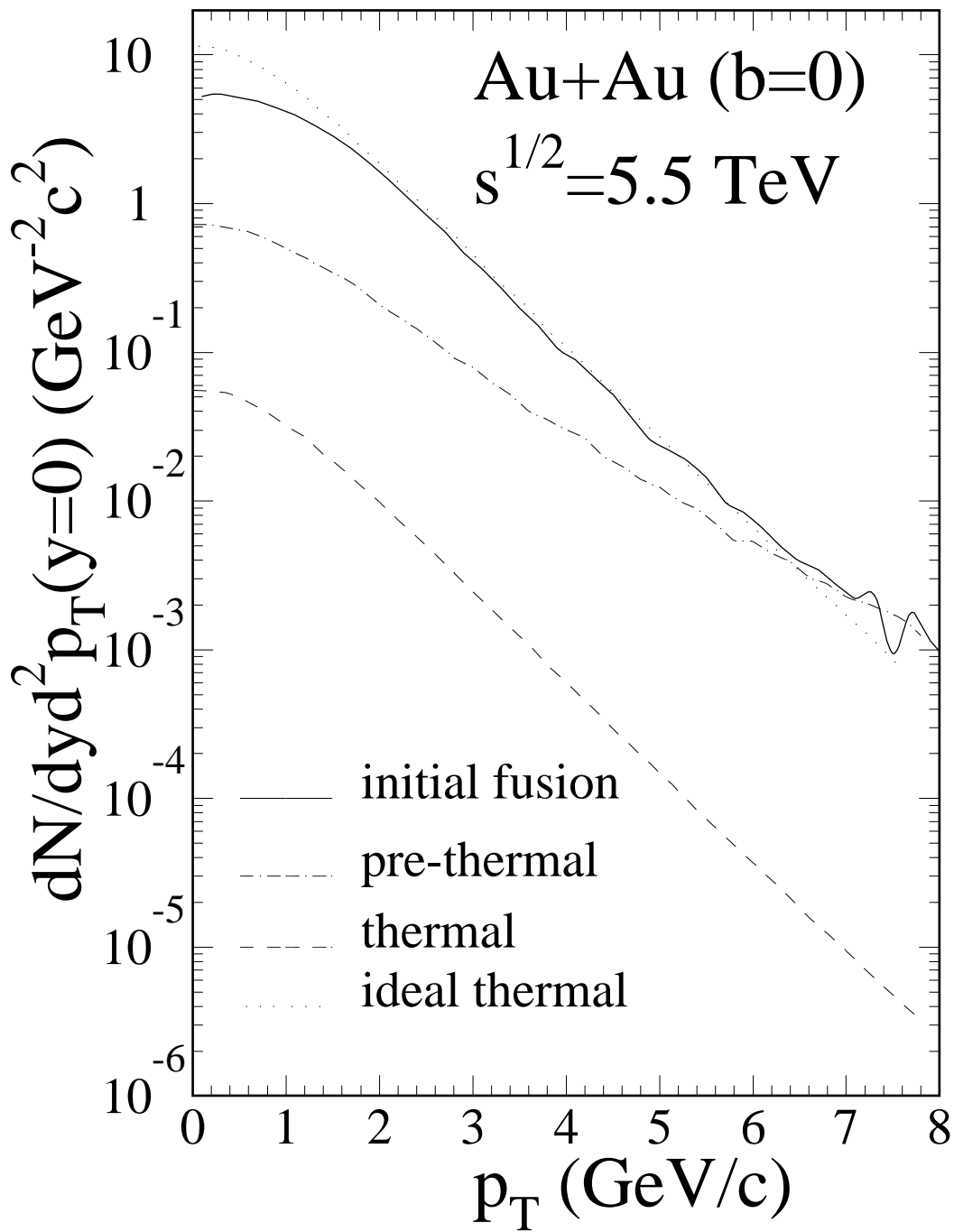
\begin{figure}
\caption{The same as in Fig.~\protect\ref{fig5}, except at LHC energy,
         $\protect\sqrt{s}=5.5$ ATeV.}
\label{fig6}
\end{figure}

During the initial interaction period, charm quarks are produced
together with minijets through gluon fusion and quark anti-quark
annihilation. Like gluon and light quark production, charm
production through the initial fusion is very sensitive to the
parton distributions inside nuclei. In addition, the cross
section is also very sensitive to the value of charm quark
mass, $M$. If higher order corrections are taken into account,
the production cross section depends also on the choices of
the renormalization and factorization scales. Detailed studies
on the next-leading-order calculation\cite{NDE,SVN,ISPV,VOGT} shows,
however, that higher order corrections to the total charm
production cross section can be accounted for by a constant
$K$-factor of about 2. This is what we will use next. For consistency
we use $M=1.5$ GeV for all calculations. Shown in Figs.~\ref{fig5}
and \ref{fig6} as solid lines are the initial charm
production given by HIJING calculations at RHIC and LHC energies,
with MRSD$-'$ \cite{MRS} parton  distributions. The corresponding total
integrated cross sections are, $\sigma_{c\bar{c}}=0.16$ (5.75) mb
at RHIC (LHC) energy, where nuclear shadowing of the gluon distribution
function is also taken into account. In HIJING calculations, high
order corrections are included via parton cascade in both initial
and final state radiations. The resultant distributions in
$c\bar{c}$-pair momentum are very close to the explicit
higher order calculations \cite{VOGT}.

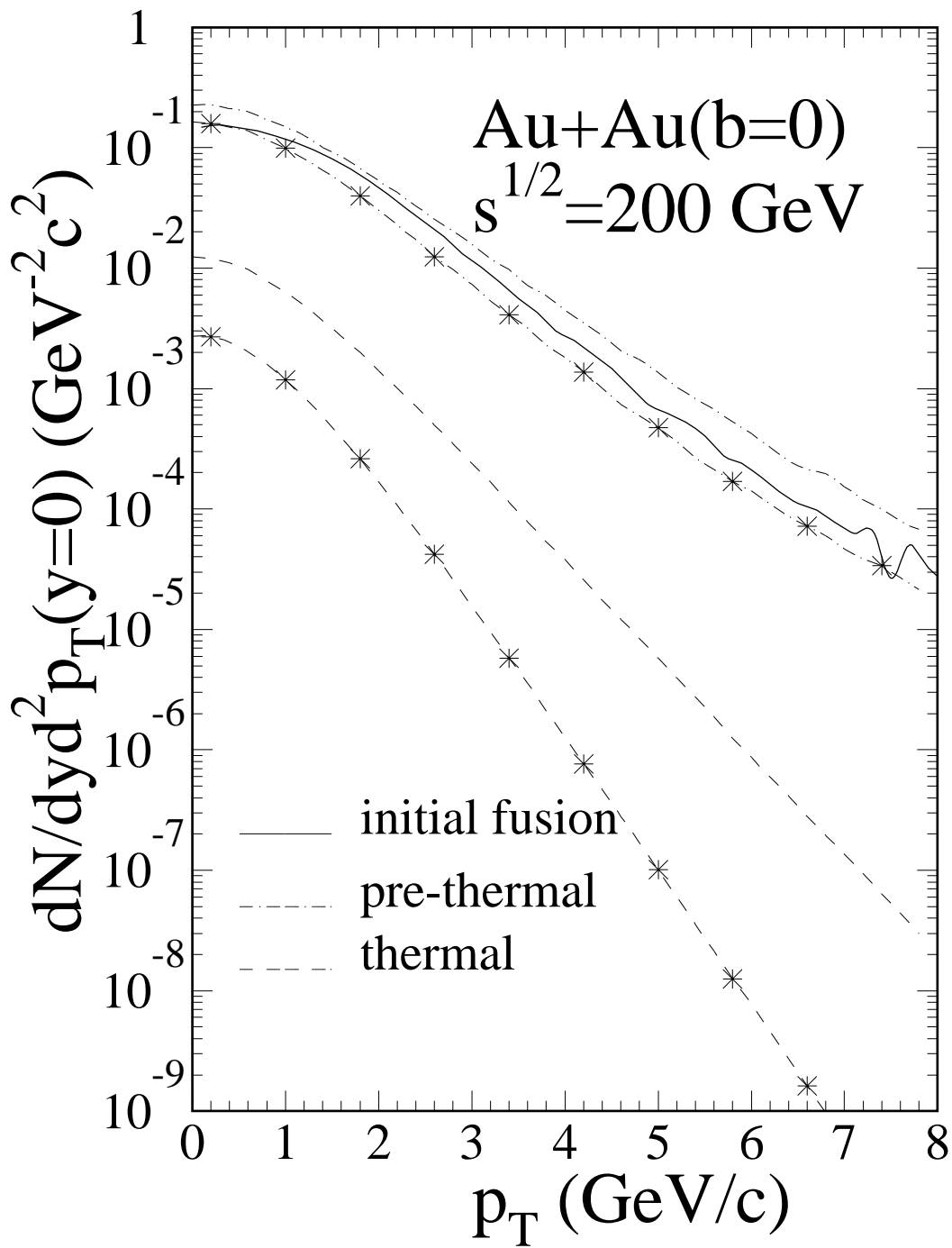
\begin{figure}
\caption{The $p_{\perp}$ distribution of the initial (solid), pre-thermal
  (dot-dashed) and thermal (dashed) charm production for
  initial parton densities 4 times higher than HIJING estimate
  given in \protect\ref{table1} but with the same (ordinary lines), or
  reduced initial temperature, $T_0=0.4$ GeV (lines with stars).}
\label{fig7}
\end{figure}

Plotted in Figs.~\ref{fig5} and \ref{fig6} as
dot-dashed and dashed lines are the pre-thermal and thermal
production. In the calculation, a factor of 2 is also
multiplied to the lowest order matrix elements of charm
production. Both contributions are
much smaller than the initial charm production at both
energies. The pre-thermal contributions shown are
also much smaller than what was found in Ref.~\cite{BMXW92}.
This is because momentum and space-time correlation
was not taken into account in Ref.~\cite{BMXW92} which
suppresses the pre-thermal charm production. Similar
results are also found in a study by Lin and Gyulassy \cite{LG94}.
As we have already discussed, the initial conditions
in Tables~\ref{table1} and \ref{table2} given by HIJING
calculations have many uncertainties. If one increases the
initial parton number density at RHIC energy by a factor
of 4 with the same initial temperature, charm production
from both pre-thermal and thermal sources will increase
about a factor of 12 as shown in Fig.~\ref{fig7},
leading to a total secondary
contribution comparable to the initial charm production.
In the extreme limit, a fully equilibrated parton plasma
($\lambda_g=\lambda_q=1$) at the same initial temperature
would give an enhancement of charm production about 4 times
higher than the initial production, shown as dotted lines
in Figs.~\ref{fig5} and \ref{fig6}. In this
case, the enhancement not only comes from higher parton
densities, but also from the much longer life time of
the parton plasma (cf. Figs.~\ref{fig2} and \ref{fig3}).
Much higher enhancements predicted in Ref.~\cite{KG93}
are due to the overestimate of the intrinsic charm production
as pointed by Gyulassy and Lin in a recent paper \cite{LG94}.
Though the intrinsic
charm production is important in the forward direction at
large $x_f$ \cite{VBH}, it is strongly suppressed in the
mid-rapidity region due to the interference among pQCD
amplitudes to the same order \cite{CSS86}.

To test the sensitivity of open charm production to
uncertainties in initial fugacities and temperature
separately, we consider an alternative scenario as
we have discussed in the parton evolution. We assume
the initial parton densities to be 4 times higher than
given in Table~\ref{table1} at RHIC energy but with lower
initial temperature, $T_0=0.4$ GeV.
Accordingly, the initial phase-space distribution
is also modified to: $h(k_{\perp})=9649.2 e^{-k_{\perp}/0.65}/(k_{\perp}+0.3)$
from the one in Table~\ref{table2}, which gives 4 times of the
initial parton density but smaller average transverse
momentum, $\langle k_{\perp}\rangle = 0.85$ GeV. The
reduced average transverse momentum corresponds to
lower initial effective temperature.
This system with higher initial
fugacities evolves faster toward equilibrium but the
life-time of the deconfined phase is shorter due to
the reduced temperature as we have discussed. The
corresponding open charm production is shown in Fig.~\ref{fig7}
by the lines with stars. We observe that open charm production
from both pre-thermal and thermal contribution is reduced due
to the reduction in initial temperature and life-time of the
parton plasma, even though the initial fugacities are much
higher and the evolution toward equilibrium is faster.
Thus, open charm production is much more sensitive to the
change in the initial temperature than the parton fugacities.
We also note from Eqs.~(\ref{eq:therm2}) and (\ref{eq:preth})
that the pre-thermal and thermal charm production depends
on the thermalization time $\tau_{\rm iso}$ and the life
time of the parton plasma. Therefore, by measuring the
charm enhancement, we can probe the initial parton phase-space
distribution, initial temperature and the thermalization
and equilibration time.

\section{Conclusions}

In this paper, we have calculated open charm production
in an equilibrating parton plasma, taking into account
the evolution of the effective temperature and parton
fugacities according to the solution of a set of rate
equations. In the evaluation of the interaction rate
$R_3$ for induced gluon radiation, a color dependent
effective formation time was used which reduces the
gluon equilibration rate through LPM suppression of
soft gluons. In the calculation of the pre-thermal
contribution to open charm production, correlation
between  momentum and space-time was also included.
This correlation reduces the pre-thermal charm production
as compared to the uncorrelated one used in a previous
estimate \cite{BMXW92}.

We found that both the thermal contribution during
the parton equilibration and pre-thermal contribution
with the current estimate of the initial parton density from
HIJING Monte Carlo simulation are much smaller than the
initial direct charm production. However, the final
total charm production is very sensitive to the initial
condition of the parton evolution. If uncertainties in
the initial parton production can increase the initial
parton density, {\em e.g.}, by a factor of 4,
the total secondary charm production
will become comparable or larger than the initial
production, due to both the increased production rate
and longer life time of the parton plasma. We also
found that open charm production is more sensitive to
the initial temperature of the parton system than the
initial parton fugacities. Therefore,
open charm production is a good probe of the initial
parton distribution in phase-space and the thermalizaion
and equilibration time of the parton plasma.

\section*{Acknowledgments}

P. L. and X.-N. W. would like to thank M. Asakawa and M. Gyulassy
for helpful  discussions. B. M. and X.-N. W. thank T. S. Bir\`o,
E. van Doorn, and M. H. Thoma for their early collaboration
in  the study of parton equilibration.
This work was supported by the Director, Office of Energy
Research, Division of Nuclear Physics of the Office of High
Energy and Nuclear Physics of the U.S. Department of Energy
under Contract No. DE-AC03-76SF00098 and DE-FG05-90ER40592.
P. L. and X.-N. W. were also supported by the U.S. - Hungary
Science and Technology Joint Fund J. F. No. 378.

\begin{table}
\begin{center}
\begin{tabular}{lrr}
&\makebox[1in]{$\;\;\;$RHIC} &\makebox[1in]{$\;\;$LHC} \hfill \\ \hline
$\tau_{\rm iso}$ (fm/$c$) &0.7\phantom{000}   &0.5\phantom{000} \hfill \\
$\varepsilon_0$ (GeV/fm$^3$)  &3.2\phantom{000}   &40\phantom{.0000} \hfill \\
$n_0$ (fm$^{-3}$)             &2.15\phantom{00}   &18\phantom{.0000}  \hfill \\
$\langle k_{\perp}\rangle$ (GeV)    &1.17\phantom{00}  &1.76\phantom{00}
\hfill\\
$T_0$ (GeV)                   &0.55\phantom{00}  &0.82\phantom{00} \hfill\\
$\lambda_g^0$                 & 0.05\phantom{00} &0.124\phantom{0} \hfill\\
$\lambda_q^0$                 &0.008\phantom{0}  &0.02\phantom{00} \hfill \\
\end{tabular}

\caption{Values of the relevant parameters characterizing the parton
plasma at the moment $\tau_{\rm iso}$, when local isotropy
of the momentum distribution is first reached.}
\label{table1}
\end{center}
\end{table}

\begin{table}
\begin{center}
\begin{tabular}{cll}
\makebox[1in]{$\sqrt{s}$ (TeV)} & $Y$ &\makebox[1in]{$h(k_{\perp})$ (GeV$^-2$)}
            \hfill \\ \hline
& &  \\
0.2 & 2.5 & $1754.4e^{-k_{\perp}/0.9}/(k_{\perp}+0.3)$ \hfill \\
& & \\
5.5 & 5.0 & $2.66\times 10^7/(k_{\perp}+2.9)^{6.4}$ \hfill \\
& & \\
\end{tabular}

\caption{Parametrizations of the momentum spectra of the initially
produced partons in HIJING calculation.}
\label{table2}
\end{center}
\end{table}

\end{document}